\documentclass[twocolumn]{aa}
\usepackage{graphicx}
\usepackage{txfonts}
\usepackage{tabularx}
\usepackage{amsmath}
\usepackage{xcolor}
\usepackage{cancel}
\usepackage{ulem}
 \usepackage{threeparttable}

\begin{document}

\title {Spectral evolution of RX~J0440.9+4431 during the  2022-2023 giant outburst observed with Insight-HXMT}
\titlerunning{Spectral evolution of RX~J0440.9+4431}

\author
{
P. P. Li \inst{1,2}
\and Peter A. Becker\inst{3}\thanks{E-mail: pbecker@gmu.edu}
\and L. Tao\inst{1}\thanks{E-mail: taolian@ihep.ac.cn}
}
% List of institutions

\institute{Key Laboratory of Particle Astrophysics, Institute of High Energy Physics, Chinese Academy of Sciences, Beijing 100049, China
\and University of Chinese Academy of Sciences, Chinese Academy of Sciences, Beijing 100049, China
\and Department of Physics and Astronomy, George Mason University, Fairfax, VA 22030-4444, USA\\
}

\authorrunning{P.P. Li et al.}
\date{Received ; accepted }

% \abstract{}{}{}{}{} 
% 5 {} token are mandatory
 
  \abstract 
   {In 2022--2023, the Be/X-ray binary X-ray pulsar RX~J0440.9+4431 underwent a Type~II giant outburst, reaching a peak luminosity $L_{\rm X} \sim 4\times10^{37}\ {\rm erg\ \rm s^{-1}}$. In this work, we utilize Insight-HXMT data to analyze the spectral evolution of RX~J0440.9+4431 during the giant outburst. By analysing the variation of the X-ray spectrum during the outburst using standard phenomenological models, we find that as the luminosity approaches the critical luminosity, the spectrum became flatter, with the photon enhancement predominantly concentrated around \textasciitilde 2\,keV and 20--40\,keV. The same behavior has also been noted in Type~II outbursts from other sources. While the phenomenological models provide good fits to the spectrum, this approach is sometimes difficult to translate into direct insight into the details of the fundamental accretion physics. Hence we have also analyzed spectra obtained during high and low phases of the outburst using a new, recently-developed physics-based theoretical model, which allows us to study the variations of physical parameters such as temperature, density, and magnetic field strength during the outburst. Application of the theoretical model reveals that the observed spectrum is dominated by Comptonized bremsstrahlung emission emitted from the column walls in both the high and low states. We show that the spectral flattening observed at high luminosities results from  a decrease in the electron temperature, combined with a compactification of the emission zone, which reduces the efficiency of bulk Comptonization. We also demonstrate that when the source is at maximum luminosity, the spectrum tends to harden around the peak of the pulse profile, and we discuss possible theoretical explanations for this behavior. We argue that the totality of the behavior in this source can be explained if the accretion column is in a quasi-critical state when at the maximum luminosity observed during the outburst.}


   \maketitle
%
%________________________________________________________________

\section{Introduction}
X-ray Pulsars (XRPs) are a class of accreting, strongly magnetized ($B \gtrsim10^{12}\ \rm G$) neutron stars (NSs) located in binary systems, and typically accompanied by massive optical companions \citep[see reviews by][]{2006Harding,2022Mushtukov}. Plasma material from the optical companion can be accreted onto the compact object through various mechanisms, including stellar winds, accretion disks, or some combination of processes. When the accreting material approaches the NS at a certain distance, its flow is disrupted by the strong magnetic field, forming a boundary at the NS magnetosphere \citep{2015Belenkaya}. Due to various instabilities near the magnetosphere radius \citep{1993Spruit}, material enters the magnetosphere and travels along magnetic field lines to reach a small region on the stellar surface surrounding the magnetic pole, with radius \textasciitilde$10^{4-5}\ \rm cm$ \citep[e.g.,][]{HardingEtal1984}. The kinetic energy of the accreting gas (derived from its gravitational potential energy) is ultimately radiated away in the form of X-rays \citep[][]{1976Basko}.

The observed luminosities of XRPs span an extremely wide range, covering \textasciitilde 9 orders of magnitude, from $L_{\rm X} \sim 10^{32}\ {\rm erg\ \rm s^{-1}}$ to $L_{\rm X} \sim 10^{41}\ {\rm erg\ \rm s^{-1}}$ \citep[e.g.][]{2017Rouco,2021Lutovinov,2020Kong}. In the low-luminosity sources, the accreting matter passes through a gas-mediated, discontinuous shock at the top of the accretion column, with the subsequent final deceleration occurring via Coulomb collisions near the stellar surface \citep{LangerAndRappaport1982}. Conversely, in the high-luminosity sources, the accreting material passes through an extended, radiation-dominated shock. In this situation, the gas decelerates smoothly through the sonic point under the influence of radiation pressure \citep{1981Wang,Davidson1973}.

The spectra of XRPs have typically been fitted using products of power-laws and exponential cutoffs \citep[e.g.,][]{WhiteEtal1983}. The exponential cutoff is interpreted physically as the onset of collisional Compton recoil losses for photons with energy $\epsilon \gtrsim kT_{e}$ scattering off electrons with temperature $T_{e}$. This interpretation has motivated the wide adoption of the {\tt cutoffpl} model in {\tt XSPEC}\footnote{\url{https://heasarc.gsfc.nasa.gov/docs/xanadu/xspec/manual/manual.html}} for researchers fitting the spectral continua of XRPs. Subsequently, other similar simplified models have been used to fit the continuum emission from XRPs, such as the models {\tt NPEX} \citep{1995Mihara} and  {\tt powerlaw} $\times$ {\tt highecut} \citep{2002Coburn}, and so on.

In the fitting of XRP spectra, one or more extra blackbody components ({\tt bbodyrad}) are sometimes needed to account for the soft emission in the spectra. Under low-luminosity conditions, typically, only one {\tt bbodyrad} needs to be included in the fitting. This is often interpreted as the cooling of an accretion-heated NS crust in the propeller state \citep{2016Tsygankov}, or as thermal radiation originating from the hot spot at the base of the accretion column \citep{2016Tsygankovb,2021AGorban}. However, in the case of luminous XRPs ($L_{\rm X} \gtrsim10^{38}\ {\rm erg\ \rm s^{-1}}$), such as 1A~0535+262 \citep{2021Kong} and SXP~59 \citep{2019Weng}, two blackbody components are often needed - one with a high temperature and another with a low temperature. The high-temperature blackbody, reaching up to 4\,keV with a radius of \textasciitilde 1\,km, is often linked to the accretion column. The lower-temperature blackbody is considered to be associated with the accretion disk or accretion curtain. Nevertheless, these methods are still relatively phenomenological and therefore it is challenging to relate them to the detailed physical processes occurring in XRP accretion columns. Hence, most current research primarily focuses on the evolution of spectral parameters \citep{2020Kong,2022Tamang,2023Mandal}. There are also clear indications of cyclotron features and iron emission lines in the observed spectra \citep[e.g.][]{2019Jaisawal,2023Tamang}.

Due to the extremely strong magnetic fields and gravity, as well as the high-temperature environment in XRPs, the radiative transfer in the accreting gas is quite complex, and it has therefore proven very challenging to simulate the physical processes occurring in these sources using theoretical models. Currently, there is no widely accepted, fully self-consistent physical model to describe the observed spectra of these objects across a wide range of mass accretion rates, because of the strong coupling between the radiative transfer and the radiation hydrodynamics that determines the physical structure of the accretion column. However, significant progress on the development of theoretical models for the calculation of XRP spectra has been accomplished in recent years, which we briefly review below.

In the case of the high-luminosity XRPs \citet[][hereafter {\tt BW07}]{BeckerWolff2007} incorporated both bulk and thermal Comptonization in their theoretical model describing the X-ray spectral formation process in XRPs. They presented an analytical derivation of the Green's function for the problem and proceeded to convolve the Green's function with source terms corresponding to bremsstrahlung, cyclotron, and blackbody emission, to compute emergent spectra. This approach was successfully applied to explain the phase-averaged X-ray spectra for sources such as Her X-1, Cen X-3, and LMC X-4. Later, \cite{2012Farinelli} adjusted the velocity profile of the accreting material based on the {\tt BW07} model and developed a {\tt compmag} model for X-ray spectral fitting software package {\tt XSPEC}. However, these models have limitations when fitting data from low-luminosity sources, because they only consider photon escape through the walls of the accretion column.

Recently, \cite[][hereafter {\tt BW22}]{BeckerWolff2022} have expanded upon {\tt BW07}, addressing aspects such as the geometry of the accretion column, the calculation of separate top and wall components for the escaping radiation beam, and incorporating a realistic velocity profile. In particular, their model can be used to analyze spectra from X-ray pulsars over a very wide range of luminosity, which they demonstrated by successfully fitting the spectra of the high-luminosity source Her X-1 and also the low-luminosity source X Per using the same model. However, further scrutiny through additional spectral data is still required to provide a solid theoretical foundation for understanding the formation of X-ray spectra emitted by XRPs. Because the model of {\tt BW22} can accommodate a wide range of accretion rates, it is ideally suited for the analysis of the spectral evolution observed during Type~II X-ray outbursts from XRPs. Hence one of the primary goals of this paper is to use the new model of {\tt BW22} to analyze the observations obtained during two portions of the giant outburst from RX~J0440.9+4431 between January and March in 2023, in which the source luminosity exhibited significant variation.

\section{Source Properties}
\label{sec:source}

RX~J0440.9+4431 is a Be/X-ray binary system (BeXRB; see review by \cite{2011Reig}). In 2022--2023, RX~J0440.9+4431 underwent a first-ever giant outburst \citep{2022Nakajima,2023ATelpal,2023ATelColey}, reaching a peak luminosity $L_{\rm X} \sim 4\times10^{37}\ {\rm erg\ \rm s^{-1}}$. Prior to the giant outburst, the system had undergone three smaller outbursts in March 2010 \citep{2010ATelMorii}, September 2010 \citep{2010ATelKrivonos}, and January 2011 \citep{2011ATelTsygankov}, but the luminosity during these events was $L_{\rm X} <10^{37}\ {\rm erg\ \rm s^{-1}}$. The X-ray spectrum of RX~J0440.9+4431 has generally been characterized by fitting the data using typical phenomenological models, based on simple mathematical functions such as power laws and exponential cutoffs. For example, \cite{2012Usui} analyzed the phase-averaged spectrum of the first outburst of RX~J0440.9+4431 observed by RXTE. They found that the spectrum was well fitted using a power-law model modified with an exponential cut-off, plus a blackbody component with a temperature of \textasciitilde 2\,keV, and an additional Gaussian function representing an iron emission line. Subsequently, \cite{2012Tsygankov} conducted an analysis of the spectrum of RX~J0440.9+4431 obtained during the September 2010 outburst using data from INTEGRAL, Swift, and RXTE. They reported the discovery of a \textasciitilde 32\,keV cyclotron resonant scattering feature (CRSF). However, \cite{2013Ferrigno} reanalyzed the data and explored different models ({\tt bmc} and {\tt compmag}). Their conclusion was that the absorption feature reported by \cite{2012Tsygankov} was likely a result of systematic biases in the joint analysis of data from multiple instruments and/or an inadequate modeling of the continuum emission.

Simple phenomenological models are often difficult to translate into physical variables such as temperature, density, and magnetic field strength. This has motivated attempts to gain additional insight via the fitting of physics-based models. For example, \cite{2023Salganik} conducted a detailed study of the spectra based on observations from NuSTAR, INTEGRAL, and NICER. The broadband energy spectrum of RX~J0440.9+4431 in its bright state was well described by a two-hump model ({\tt comptt}+{\tt comptt}) with peaks at energies around 10--20\,keV and 50--70\,keV, without any observed photon deficit around 30\,keV. The high-energy component in it may be caused by multiple Compton scattering of cyclotron seed photons in the neutron star atmosphere. Additionally, \cite{2023Mandal} used NICER data to analyze the evolution of continuum spectral parameters and iron line parameters for this source.

In this work, we utilize Insight-HXMT data, which covers a broader energy range than NICER, to analyze the spectral evolution of RX~J0440.9+4431 during the giant outburst in 2022--2023. The observations and data reduction procedures are presented in Section~\ref{sec:2}. Based on phenomenological fits, we find that as the luminosity approached the critical luminosity, the spectrum became flatter, with photon enhancement predominantly concentrated around \textasciitilde 2\,keV and 20--40\,keV. In order to gain additional physical insight, beyond what can be obtained using the phenomenological models, in Section~\ref{sec:3} we analyze the spectral data obtained during high and low luminosity phases of the giant outburst using the physics-based {\tt BW22} model. Finally, in Section~\ref{sec:4} we discuss our principle findings and conclusions.

%%%%%%%%%%%%%%%%%%%%%%%%%%%%%%%%%%%%%%%%%%%%%%%%%%
\begin{figure*}
\centering
\includegraphics[width=0.48\textwidth]{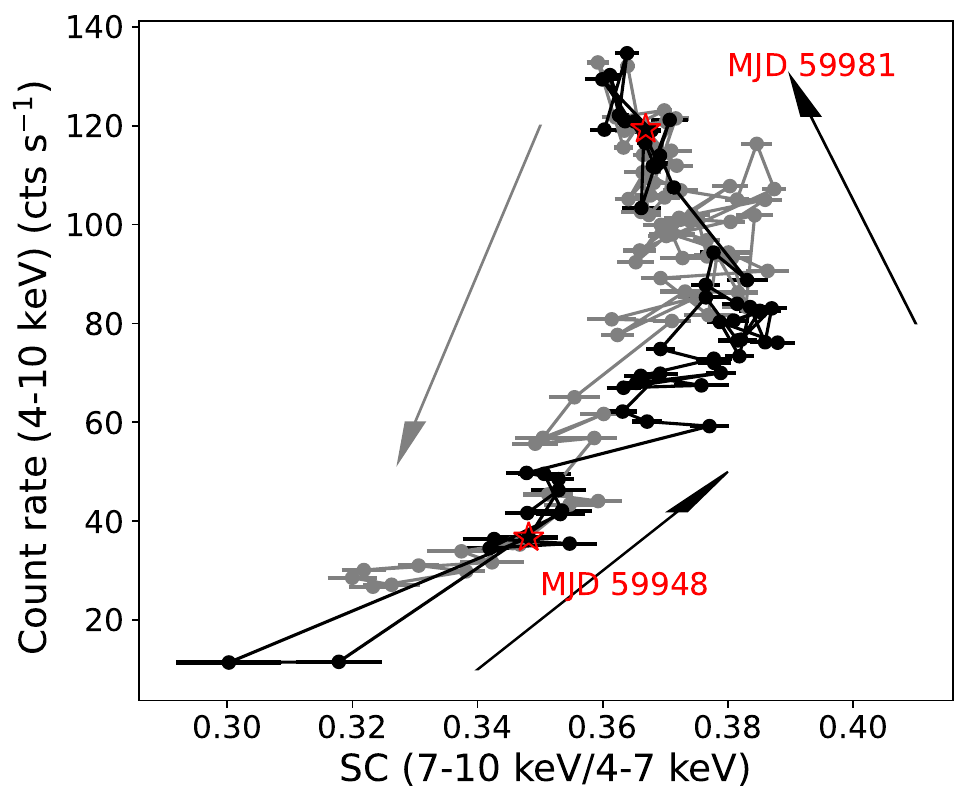}
\includegraphics[width=0.48\textwidth]{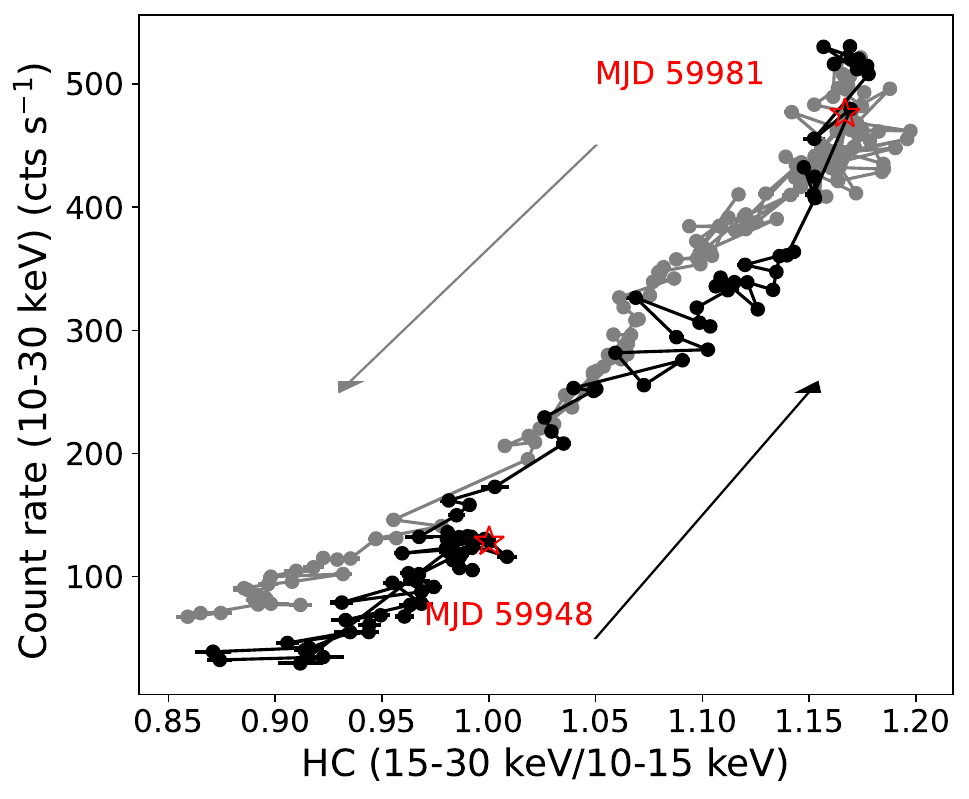}
\caption{The hardness-intensity diagram (HID) of RX J0440.9+4431 in 2022--2023 giant outburst observed with Insight-HXMT. The arrows describe the direction of evolution. The black and gray colors represent the rising and declining parts of the outburst, respectively. The red stars indicate the representative low and high state data used for the spectral fits.}
  \label{fig:4}
\end{figure*}

%%%%%%%%%%%%%%%%%%%%%%%%%%%%%%%%%%%%%%%%%%%%%%%%%%%%%%%%%%%%%%

\section{Observations and Data Reduction}
\label{sec:2}

Insight-HXMT is China's first space astronomical satellite, equipped with three payloads capable of covering a total energy range of 1--250\,keV \citep{zhang2020}. These payloads include Low Energy (LE) detectors (1--15\,keV) with an area of 384 ${\rm cm}^2$ \citep{2020SCPMAChen}, Medium Energy (ME) detectors (5--30\,keV) with an area of 952 ${\rm cm}^2$ \citep{2020SCPMACao}, and High Energy (HE) detectors (20--250\,keV) with an area of 5100 ${\rm cm}^2$ \citep{2020SCPMALiu}. Since its launch in 2017, many interesting results regarding accreting pulsars have been reported \citep[e.g.][]{2022Wang0535,2022Hou,2023Fu}.

Following the X-ray outburst of RX~J0440.9+4431, which was triggered by MAXI, Insight-HXMT conducted intensive observations for nearly three months, from MJD 59944 to 60029. These observations covered various phases of the outburst, including the rise, the outburst peak, and the full decay phase. The ObsIDs used in this study have been provided in the article on the timing analysis of RX~J0440.9+4431 \citep{2023Li}. All of the observations studied here were analyzed using the Insight-HXMT processing software {\tt HXMTDAS v2.05}\footnote{\url{http://hxmtweb.ihep.ac.cn/software.jhtml}}, following standard processing procedures\footnote{\url{http://hxmtweb.ihep.ac.cn/SoftDoc/648.jhtml}}. The energy ranges selected were 2--10\,keV (LE), 10--30\,keV (ME), and 30--120\,keV (HE). To improve the counting statistics of the spectra, we utilized the ftool {\tt grppha}. For the LE data, channels 0--579 are binned up by a factor of 2, and channels 580--1535 are binned up by a factor of 10. For the ME data, channels 0--1024 are binned up by a factor of 2. Similarly, for the HE data, channels 0--255 are binned up by a factor of 2. The spectral data were analyzed using {\tt XSPEC v12.12.1}, with a system error set to 1\% for the HE range.

%%%%%%%%%%%%%%%%%%%%%%%%%%%%%%%%%%%%%%%%%%%%%%%%%%
\begin{figure*}
\centering
\includegraphics[width=0.48\textwidth]{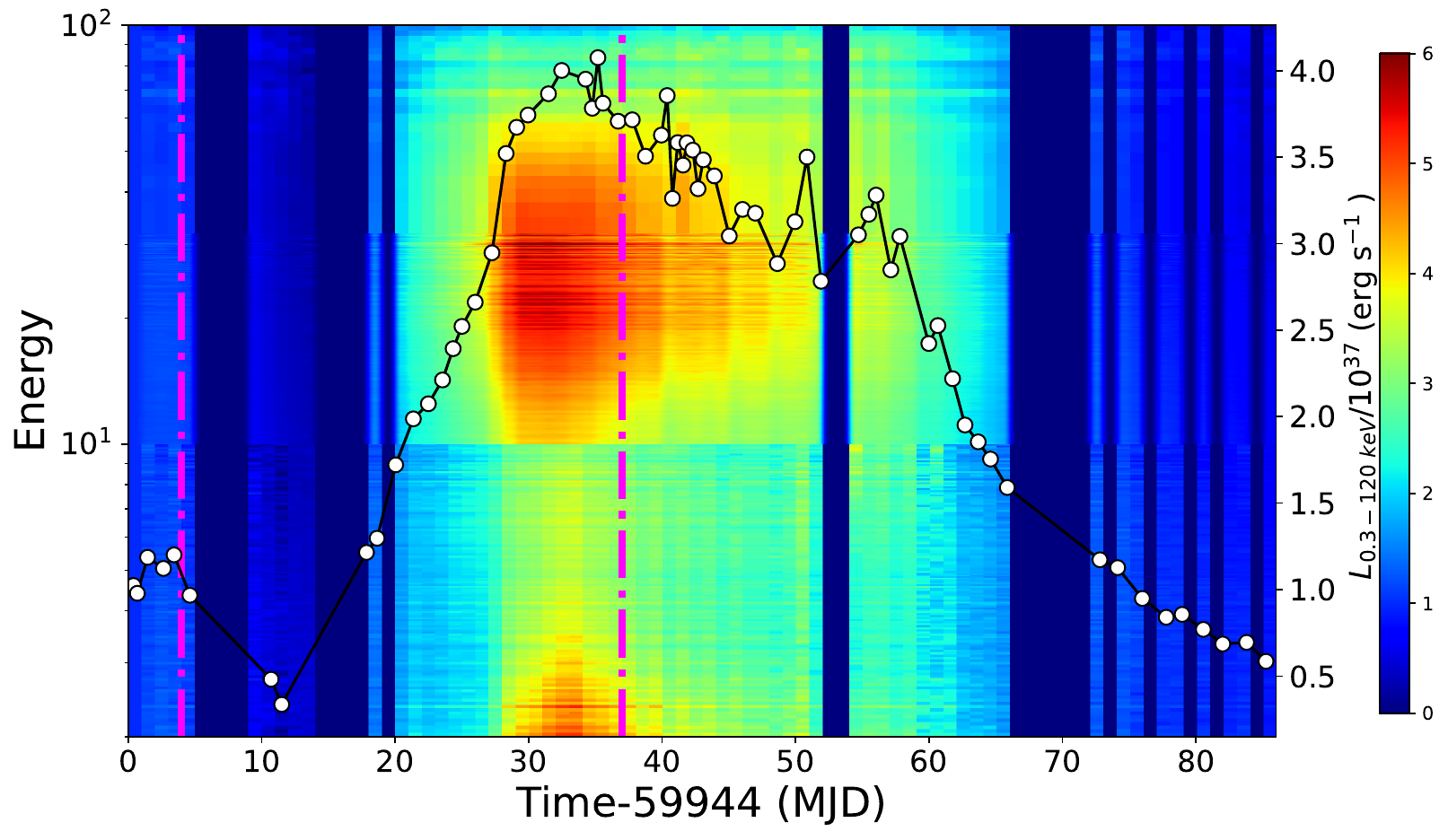}
\includegraphics[width=0.48\textwidth]{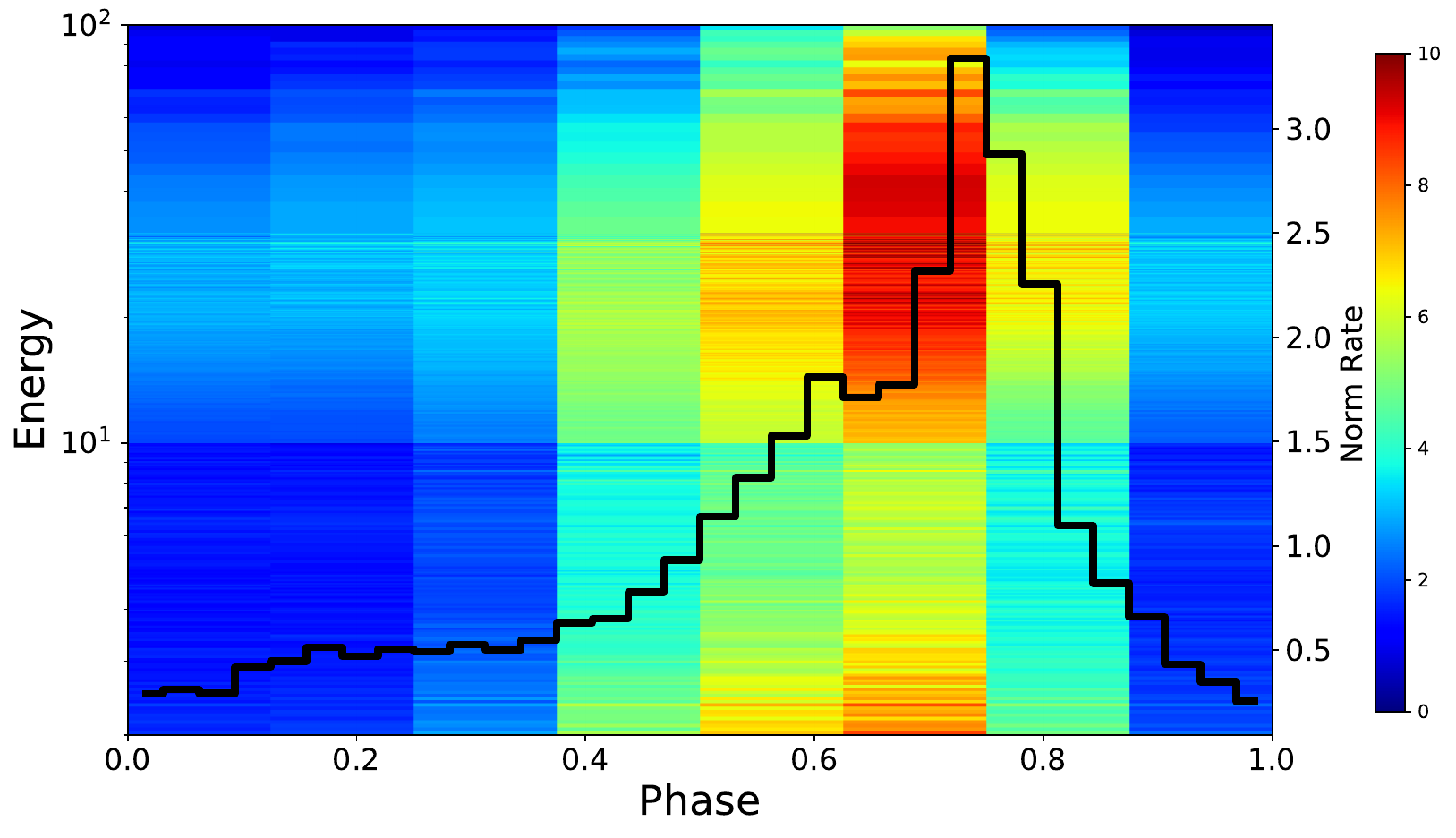}
\caption{Left panel: Spectral evolution and pulse profile of RX~J0440.9+4431 during the outburst, based on Insight-HXMT data. The color bar signifies flux intensity, with dark blue denoting periods without observations. White dots represent X-ray luminosity in the 0.3--120\,keV energy range, aligned with the right y-axis. Two dashed dotted lines (MJD 59948; MJD 59981) indicate the positions of ObsID P0514361005 (low state) and ObsID P0514361040 (high state), respectively. Right panel: The evolution of the energy ratio over phase comes from ObsID P0514361040. The solid black line represents the pulse profile for 70--100\,keV \citep{2023Li}, corresponding to the right y-axis.}
  \label{fig:1}
\end{figure*}

%%%%%%%%%%%%%%%%%%%%%%%%%%%%%%%%%%%%%%%%%%%%%%%%%%%%%%%%%%%%%%

%%%%%%%%%%%%%%%%%%%%%%%%%%%%%%%%%%%%%%%%%%%%%%%%%%%%%%%%%%%%%
\begin{figure*}
 \centering
	\includegraphics[width=0.8\textwidth]{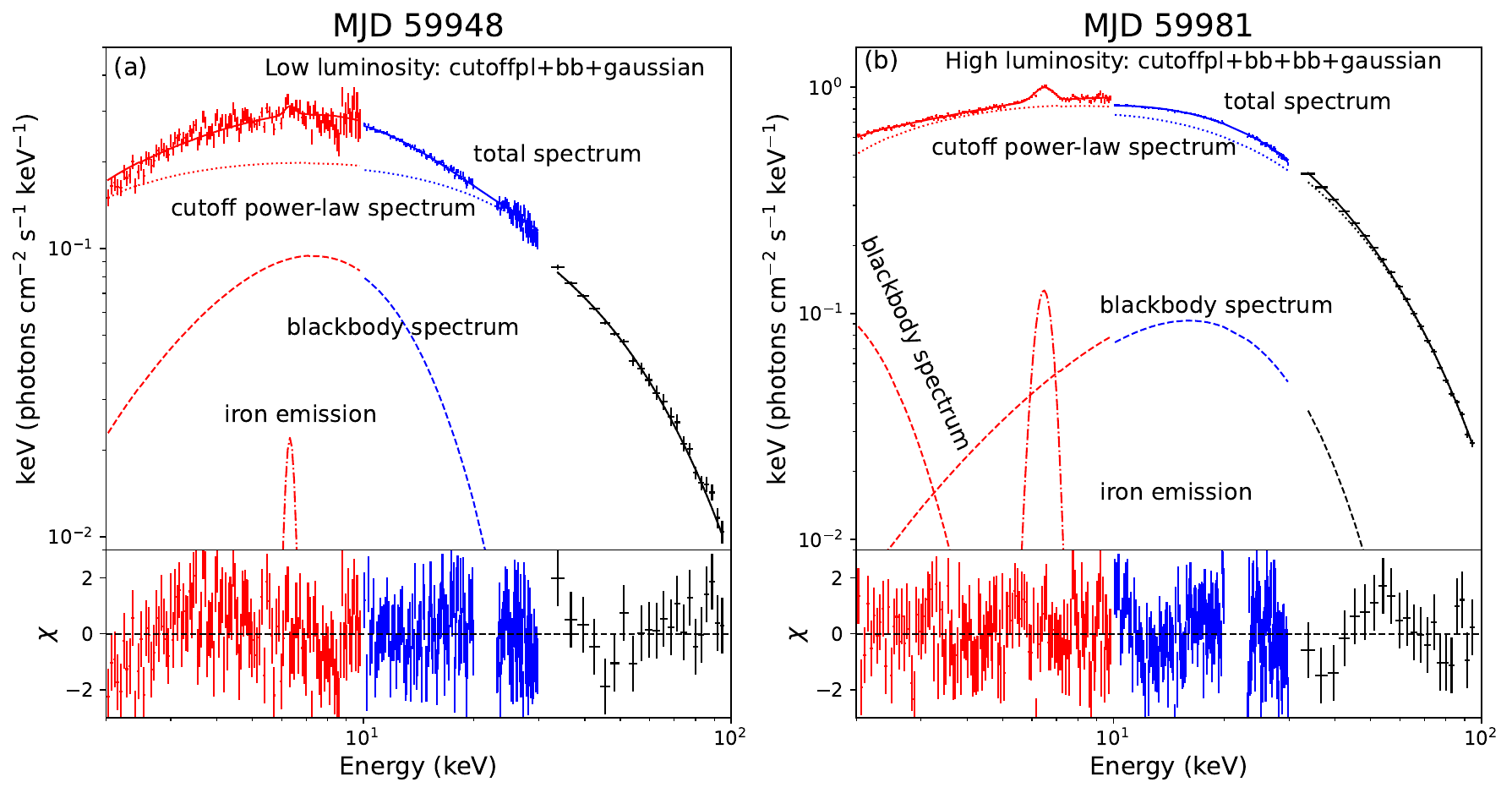}
    \includegraphics[width=0.8\textwidth]{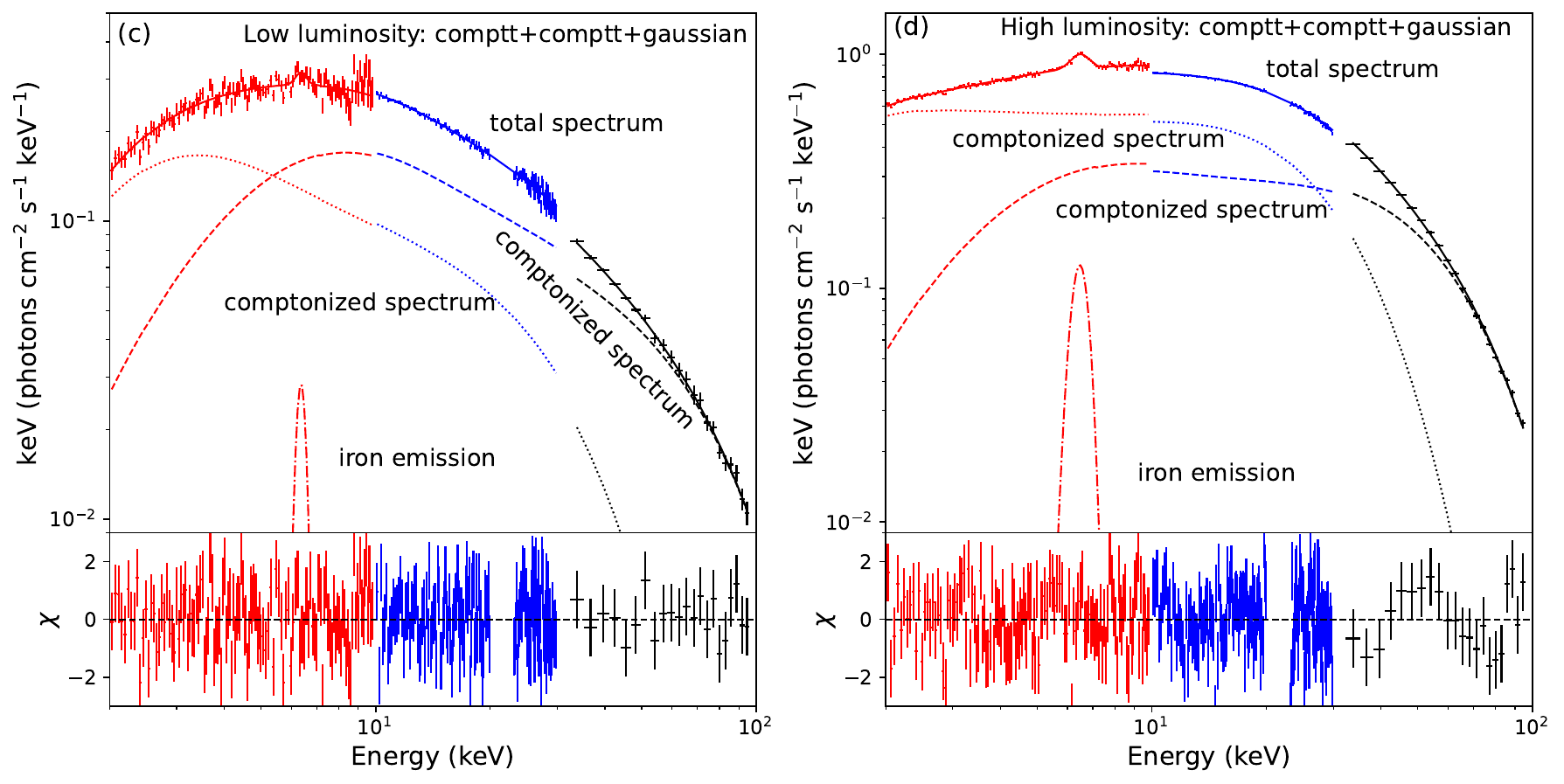}
    \caption{Unfolded spectra, phenomenological model components, and spectral residuals for Insight-HXMT observations (red for LE; blue for ME; black for HE) of RX~J0440.9+4431 with {\tt model 1} and {\tt model 2}.}
  \label{fig:2}
\end{figure*}

%%%%%%%%%%%%%%%%%%%%%%%%%%%%%%%%%%%%%%%%%%%%%%%%%%%%%%%%%%%%

%%%%%%%%%%%%%%%%%%%%%%%%%%%%%%%%%%%%%%%%%%
\begin{table*}
\centering
\caption{Best-fitting spectral parameters of the Insight-HXMT observations of RX~J0440.9+4431 using {\tt model 1} and {\tt model 2}. Uncertainties are given at 90\%.} 
\renewcommand\arraystretch{1.6}
\setlength{\tabcolsep}{2.8mm}{
\label{tab:obsid_1}
\begin{tabular}{cccccc}

\hline
ObsID      &            & \multicolumn{2}{c}{P0514361005}(low state) & \multicolumn{2}{c}{P0514361040}(high state) \\ \cline{3-6} 
MJD        &            & \multicolumn{2}{c}{59948}       & \multicolumn{2}{c}{59981}       \\ \cline{3-6} 
Components & Parameters & Model 1        & Model 2        & Model 1        & Model 2        \\ \hline
TBabs      & $N_{\rm H}\ (\times10^{22}\ {\rm cm^{-2}})$    &0.6 (fixed)   & 0.6 (fixed)                & 0.6 (fixed)              & 0.6 (fixed) \\
gaussian   & $E_{\rm Fe}\ ({\rm keV})$      & $6.32^{+0.22}_{-0.21}$       &$6.36^{+0.22}_{-0.21}$      & $6.45^{+0.03}_{-0.03}$   &$6.45^{+0.03}_{-0.03}$ \\
           & $\sigma\ ({\rm keV})$                       & $<0.41$                  & $<0.47$                    & $0.37^{+0.05}_{-0.05}$   &$0.35^{+0.05}_{-0.05}$  \\
           & $N_{\rm gau}\ (\times10^{-2})$ & $0.2^{+0.1}_{-0.1}$          & $0.2^{+0.2}_{-0.1}$        &  $1.8^{+0.2}_{-0.2}$     & $1.7^{+0.2}_{-0.2}$   \\
bbodyrad1  & $kT_{\rm 1}\ ({\rm keV})$      & $2.58^{+0.07}_{-0.07}$       & ~.~.~.                     & $0.37^{+0.04}_{-0.04}$   &  ~.~.~.            \\
           & $N_{\rm bb1}$                  & $3.7^{+0.4}_{-0.4}$          & ~.~.~.                    & $3225^{+3283}_{-1400}$   &  ~.~.~.             \\
bbodyrad2  & $kT_{\rm 2}\ ({\rm keV})$      &  ~.~.~.                      & ~.~.~.                     & $5.67^{+0.13}_{-0.15}$   &  ~.~.~.             \\
           & $N_{\rm bb2}$                  &  ~.~.~.                      & ~.~.~.                    & $0.39^{+0.05}_{-0.05}$   &  ~.~.~.             \\
cutoffpl   & $\Gamma$                       & $0.75^{+0.02}_{-0.02}$       & ~.~.~.                    & $0.62^{+0.02}_{-0.02}$   &  ~.~.~.             \\
           & $E_{\rm cut}\ ({\rm keV})$     & $25.89^{+0.65}_{-0.63}$      & ~.~.~.                    & $19.92^{+0.32}_{-0.31}$  &  ~.~.~.            \\
           & $N_{\rm cut}$                  & $0.16^{+0.01}_{-0.01}$       & ~.~.~.                    & $0.56^{+0.02}_{-0.02}$   &  ~.~.~.             \\
comptt1    & $T_{\rm 01}\ ({\rm keV})$      &  ~.~.~.                      & $0.72^{+0.07}_{-0.09}$     &  ~.~.~.                  & $0.30^{+0.05}_{-0.11}$   \\
           & $kT_{\rm 1}\ ({\rm keV})$      &  ~.~.~.                      & $6.59^{+1.03}_{-0.68}$     &  ~.~.~.                  & $6.19^{+0.10}_{-0.11}$   \\
           & $\tau_{\rm 1}$                 &  ~.~.~.                      & $9.97^{+1.23}_{-1.25}$     & ~.~.~.                   & $14.35^{+0.15}_{-0.16}$  \\
           & $N_{\rm comp1}$                &  ~.~.~.                      &$0.070^{+0.008}_{-0.004}$   & ~.~.~.                   & $0.452^{+0.090}_{-0.025}$ \\
comptt2    & $T_{\rm 02}\ ({\rm keV})$      &  ~.~.~.                      & $1.86^{+0.20}_{-0.20}$     &  ~.~.~.                  & $1.73^{+0.05}_{-0.05}$  \\
           & $kT_{\rm 2}\ ({\rm keV})$      &  ~.~.~.                      & $16.22^{+1.03}_{-0.74}$    &  ~.~.~.                  & $12.91^{+0.23}_{-0.22}$ \\
           & $\tau_{\rm 2}$                 &  ~.~.~.                      & $5.97^{+0.39}_{-0.44}$     & ~.~.~.                   & $8.94^{+0.30}_{-0.30}$  \\
           & $N_{\rm comp2}$                &  ~.~.~.                      & $0.029^{+0.003}_{-0.003}$  &  ~.~.~.                  & $0.087^{+0.003}_{-0.003}$ \\ \hline
           & $\chi^2$/d.o.f        & 348/286             & 279/283             & 342/284              & 317/283             \\ \hline
\end{tabular}
}
\end{table*}
%%%%%%%%%%%%%%%%%%%%%%%%%%%%%%%%%%%%%%%%%%%%

\section{Analysis and Results}
\label{sec:3}
Hardness-intensity diagrams (HIDs) are a very useful tool that allows investigating spectral changes without assuming any spectral model. Following the approach of \cite{2008Reig} and \cite{2013Reig}, we present in Fig.~\ref{fig:4} the count rate (4--10\,keV) as a function of the soft color (SC, 7-10\,keV/4-7\,keV) using Insight-HXMT/LE, and also the count rate (10--30\,keV) as a function of the hard color (HC, 15-30\,keV/10-15\,keV) using Insight-HXMT/ME. Fig.~\ref{fig:4} is consistent with the HID of RX J0440.9+4431 obtained by \cite{2023Salganik} and \cite{2023Mandal} using NICER data. The plot of the SC in the left-hand panel of Fig.~\ref{fig:4} exhibits two branches. Along the low-intensity branch, the SC is directly proportional to the count rate, but along the high-intensity branch, the SC is inversely proportional to the count rate, indicating a significant increase in low-energy photons with increasing luminosity. We note that the high-intensity branch in the SC plot is similar to the diagonal branch discussed by \cite{2013Reig}, which they interpret as indicative of a super-critical luminosity. However, the high-intensity branch in the SC plot in Fig.~\ref{fig:4} is shorter, perhaps suggesting that the luminosity is quasi-critical rather than super-critical. The HC for RX J0440.9+4431, plotted in the right-hand panel of Fig.~\ref{fig:4}, does not exhibit branches; instead, it shows that the HC is directly proportional to the count rate throughout the entire outburst.

The combination of the SC and HC variation plotted in Fig.~\ref{fig:4} suggests that a complex spectral evolution is occurring during the Type~II outburst studied here. To better visualize the spectral evolution of RX~J0440.9+4431 during the outburst, in the left-hand panel of Fig.~\ref{fig:1} we plot the spectral evolution, using the first observation (ObsID P0514361001) as a reference and normalizing the other data to this reference. We note that as the luminosity increases, the increase in the emission is mainly concentrated around \textasciitilde 2\,keV and \textasciitilde 20--40\,keV. In order to determine whether the increase in photons is phase-dependent, we accumulated the phase-resolved spectra into 8 pulse-phase bins from ObsID P0514361040, corresponding to the high state. Each phase-resolved spectrum is also normalized by the reference spectrum (ObsID P0514361001). The resulting pulse profile, plotted in the right-hand panel of Fig.~\ref{fig:1}, indicates that the observed spectrum becomes significantly harder at the peak of the pulse profile. This issue is further discussed from a theoretical perspective in Sec.~\ref{sec:4}.

\subsection{Phenomenological model}
\label{sec:3.1}

Initially, we analyze the variable spectrum of RX~J0440.9+4431 using the commonly used model, {\tt cutoffpl}+{\tt bbodyrad}, and include a {\tt gaussian} to describe the iron emission line at 6.4\,keV. Considering the interstellar absorption, we apply {\tt TBabs} to these three components. The hydrogen column density $N_{\rm H}$ is fixed at $0.6\times10^{22}\ \rm cm^{-2}$, which is estimated based on the Galactic H I density in the direction of RX~J0440.9+4431 \citep{2016HI4PI}. Abundances are set to WILM \citep[]{2000Wilms}, and cross-sections to VERN \citep[]{1996Verner}. Additionally, {\tt constant} is included to account for the normalization discrepancy between different telescopes, with the factor fixed at 1 for LE and free for ME and HE. Due to the uncertain calibration of an Ag emission line, we ignore the 20--23\,keV energy range in the ME. Taking ObsID P0514361005 as an example of the low-luminosity state, the {\tt model 1} ({\tt const}$\times${\tt TBabs}$\times${(\tt cutoffpl}+{\tt bbodyrad}+{\tt gaussian)}) provides a good fit at low luminosities ($\chi^{2}$ = 353 for 293 degrees of freedom, dof). The inclusion of blackbody components in the spectral fitting process is generally thought to reflect thermal emission from the magnetic pole hot spots, the optically thick accretion curtain, and/or the accretion disk. The detailed results are shown in Fig.~\ref{fig:2} and Tab.~\ref{tab:obsid_1}.

As the luminosity reaches the highest value observed during the outburst studied here, we find that significant spectral changes occur, as shown in Fig.~\ref{fig:1}. In particular, we require an extra {\tt bbodyrad} component in order to obtain an acceptable fit. Here, we select ObsID P0514361040 as an example of the high-luminosity state. In this state, we find that one blackbody has a lower temperature at $0.37\pm{0.04}$\,keV, with an emitting region radius of 10--19\,km calculated using $N_{\rm bb}$ = $(R_{\rm bb}/D_{\rm 10})^{2}$, where $D_{\rm 10}$ represents the source distance in units of 10\,kpc and $R_{\rm bb}$ is the radius of the blackbody. The other blackbody, however, has a temperature as high as $5.67^{+0.15}_{-0.12}$\,keV, corresponding to an emitting region of 0.14--0.16\,km. Similar features have been observed in the spectral fitting results of several sources, such as 1A 0535+262 \citep{2022Mandal},  Swift J0243.6+6124 \citep{2019Tao}, SXP 59 \citep{2019Weng}.

In the high state, the emission of a cold blackbody component with a radius of 10--19\,km may be related to the optically thick accretion curtain, while this hot blackbody component is not typical for Be/X-ray systems \citep{2004Hickox..881H}. Theoretically, due to the accretion column, we cannot directly receive emission from the hotspot at high brightness. Hence, \cite{2023Salganik} used two comptt models to fit the spectrum of RX~J0440.9+4431. Following his approach, we also employed {\tt model 2} ({\tt const}$\times${\tt TBabs}$\times$({\tt comptt}+{\tt comptt}+{\tt gaussian)}) to fit the low and high luminosity states separately, as shown in Fig.~\ref{fig:2} and Tab.~\ref{tab:obsid_1}. The optical depth in the high state is significantly increased compared to that in the low state, which results in a flatter spectrum.
Both {\tt model 2} and {\tt model 1} well describe the spectra, as anticipated due to their numerous parameters offering flexibility.
Yet, these parameters fall short of elucidating the fundamental physical processes.

%%%%%%%%%%%%%%%%%%%%%%%%%%%%%%%%%%%%%%%%%%%%%%%%%%%%%%%%%%%%%
\begin{figure*}
%\centering
%\begin{subfigure}{.5\textwidth}
  \centering
  \includegraphics[width=0.48\linewidth]{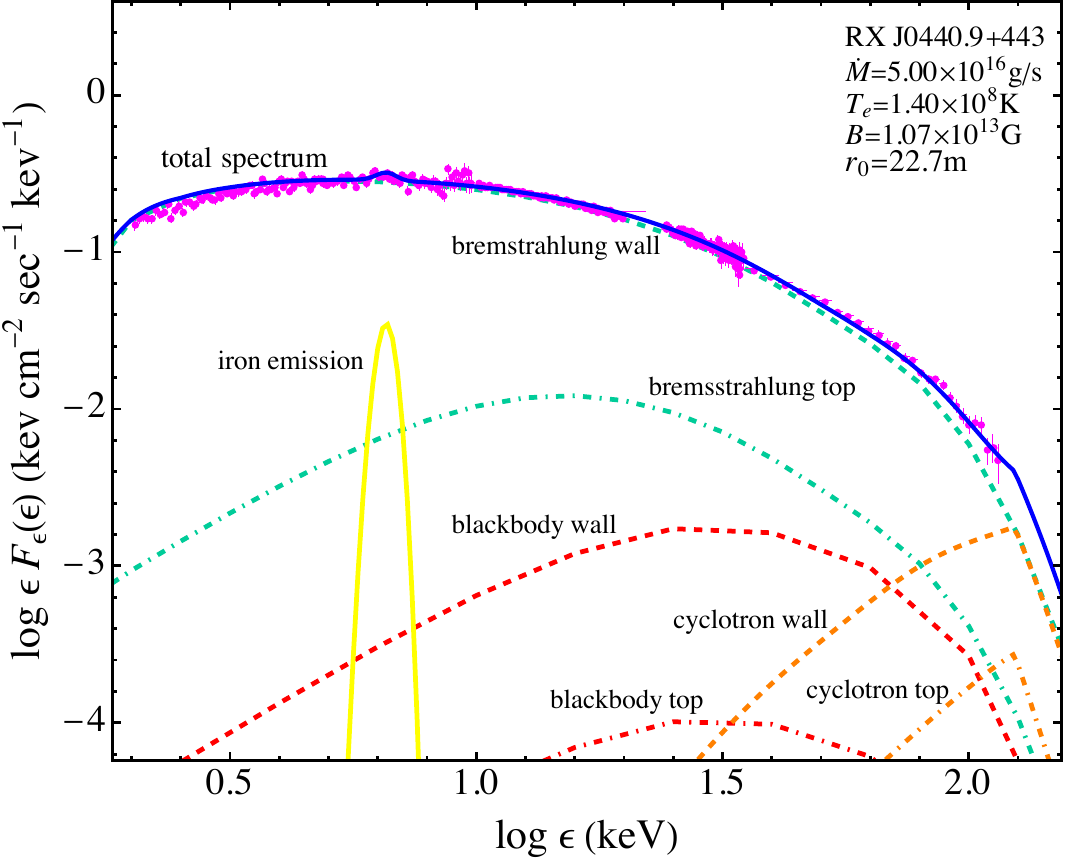}
  \includegraphics[width=0.48\linewidth]{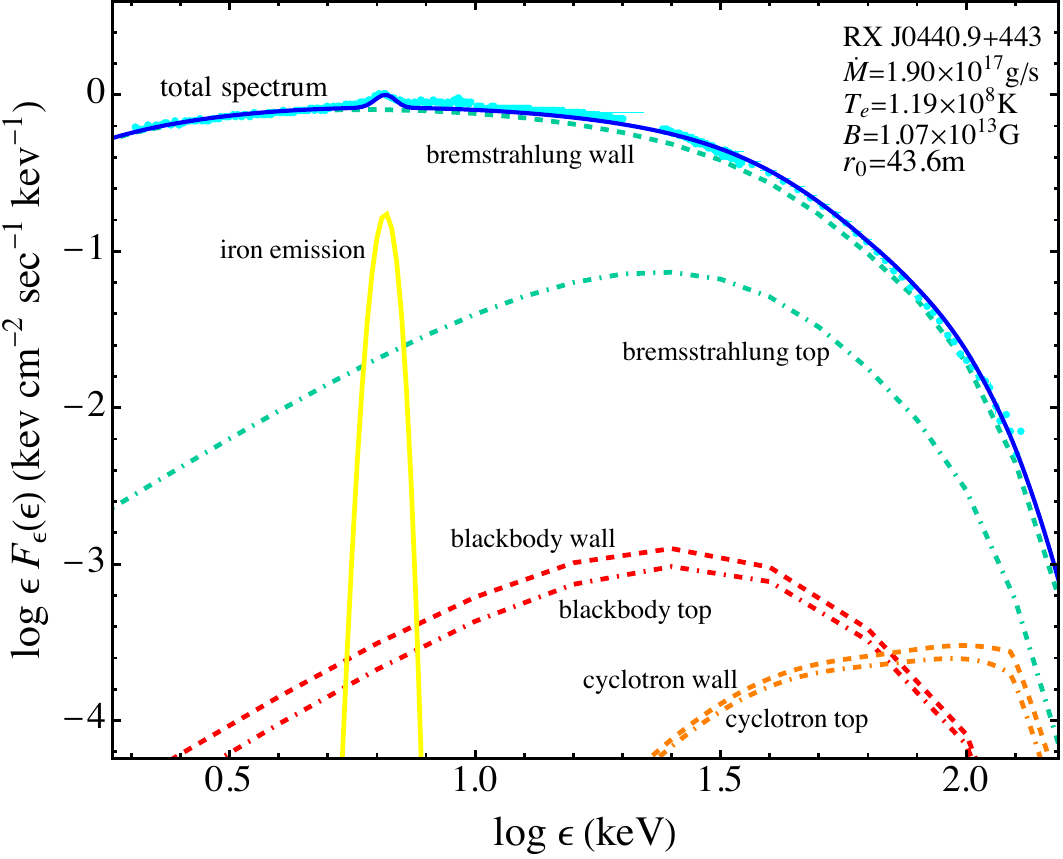}
\caption{Theoretical X-ray spectrum computed using the model of {\tt BW22}, compared with Insight-HXMT data for RX~J0440.9+443. The cyan points in the left panel refer to the low state (MJD~59948), with luminosity $L_{\rm X} = 9.23 \times 10^{36}\,{\rm erg\,s}^{-1}$, and the magenta points in the right panel refer to the high state (MJD~59981), with luminosity $L_{\rm X} = 3.44 \times 10^{37}\,{\rm erg\,s}^{-1}$. The theoretical spectral components computed using the {\tt BW22} model, corresponding to the reprocessing of bremsstrahlung, cyclotron, and blackbody seed photons escaping through the walls and top of the conical accretion column are indicated. The solid blue curves denote the total spectra.}
\label{fig:BWspec}
\end{figure*}
%%%%%%%%%%%%%%%%%%%%%%%%%%%%%%%%%%%%%%%%%%%%%%%%%%%%%%%%%%%%

%%%%%%%%%%%%%%%%%%%%%%%%%%%%%%%%%%%%%%%%%%
\begin{table*}
\centering
\caption{Input theoretical parameters used for qualitatively fitting the {\tt BW22} model to the Insight-HXMT data for high and low states.} 
\renewcommand\arraystretch{1.6}
\setlength{\tabcolsep}{2.0mm}{
\label{tab:modpar}
\begin{tabular}{ccccccccccccc}
\hline
State &$\dot M\,({\rm g\,s^{-1}})$ &$T_e\,({\rm K})$ &$B_*\,({\rm G})$ &$D\,({\rm kpc})$&$\alpha$ &$\xi$ &$\psi$ &$k_0$ &$k_\infty$  &$\Theta_1$ &$\Theta_2$ &$r_{\rm top}/R_*$              \\
\hline
high   &$1.90 \times 10^{17}$ &$1.19 \times 10^8$ &$1.07 \times 10^{13}$ &2.4   &1.20 &1.55 &1.20 &0.0 &1.0 &$0.25^\circ$ &$0.0^\circ$ &3.1

\\
low &$5.00 \times 10^{16}$ &$1.40 \times 10^8$ &$1.07 \times 10^{13}$ &2.4  &0.80 &2.47 &1.50 &0.1 &1.0 &$0.13^\circ$ &$0.0^\circ$ &3.1

\\
\hline
\end{tabular}
}
\end{table*}
%%%%%%%%%%%%%%%%%%%%%%%%%%%%%%%%%%%%%%%%%%%%

%%%%%%%%%%%%%%%%%%%%%%%%%%%%%%%%%%%%%%%%%%
\begin{table*}
\centering
\caption{Computed auxiliary theoretical parameters for {\tt BW22} model.} 
\renewcommand\arraystretch{1.6}
\setlength{\tabcolsep}{2.0mm}{
\label{tab:modpar2}
\begin{tabular}{ccccccccccccccc}

\hline
State &$\sigma_\perp/\sigma_{_{\rm T}}$ &$\sigma_{_{||}}/\sigma_{_{\rm T}}$ &$\bar\sigma/\sigma_{_{\rm T}}$ &$r_0\,({\rm m})$ &$z_{\rm th}\,({\rm m})$ &$T_{\rm th}\,({\rm K})$ &$\upsilon_*/c$ &$\upsilon_{\rm th}/c$ &$\tau_{\rm th}$ &$\tau_{\rm top}$  
%&$B_c\,({\rm G})$ &$\epsilon_c$ &$\sigma_c$ &$d_c$
\\
\hline
high      &1.00 &$1.23 \times 10^{-5}$ &$2.04 \times 10^{-4}$ &43.6   &8.07 &$7.08 \times 10^7$ &0.0 &0.03 &0.026 &1.04
\\
low   &1.00 &$1.89 \times 10^{-5}$ &$1.42 \times 10^{-4}$ &22.7   &0.0 &$1.00 \times 10^8$ &0.064 &0.064 &0.0 &1.48
\\
\hline
\end{tabular}
}
\end{table*}
%%%%%%%%%%%%%%%%%%%%%%%%%%%%%%%%%%%%%%%%%%%%

%%%%%%%%%%%%%%%%%%%%%%%%%%%%%%%%%%%%%%%%%%
\begin{table*}
\centering
\caption{Cyclotron absorption parameters for {\tt BW22} model.} 
\renewcommand\arraystretch{1.6}
\setlength{\tabcolsep}{2.0mm}{
\label{tab:modpar3}
\begin{tabular}{ccccccccccccccc}

\hline
State &$r_c/R_*$ &$B_c\,$(G) &$\epsilon_c\,$(keV)
&$\sigma_c\,$(keV) &$d_c\,$(keV) \\
\hline
high      &1.27 &$5.19 \times 10^{12}$ &60.0 &33.0   &40.3
\\
low   &1.35 &$4.33 \times 10^{12}$ &50.1 &20.0   &12.0 \\
\hline
\end{tabular}
}
\end{table*}
%%%%%%%%%%%%%%%%%%%%%%%%%%%%%%%%%%%%%%%%%%%%

%%%%%%%%%%%%%%%%%%%%%%%%%%%%%%%%%%%%%%%%%%
\begin{table*}
\centering
\caption{Distribution of radiated power over {\tt BW22} model emission components in high and low states.}
\renewcommand\arraystretch{1.6}
\setlength{\tabcolsep}{2.0mm}{
\label{tab:lumdist}
\begin{tabular}{ccccccccccccccc}

\hline
State &$L_{\rm brem}^{\rm wall}$ &$L_{\rm brem}^{\rm top}$ &$L_{\rm bb}^{\rm wall}$
&$L_{\rm bb}^{\rm top}$ &$L_{\rm cyc}^{\rm wall}$ &$L_{\rm cyc}^{\rm top}$ \\
\hline
high      &$8.9 \times 10^{-1}$ &$1.0 \times 10^{-1}$ &$1.9 \times 10^{-3}$ &$1.5 \times 10^{-3}$   &$9.4 \times 10^{-4}$ &$7.7 \times 10^{-4}$
\\
low   &$9.2 \times 10^{-1}$ &$5.1 \times 10^{-2}$ &$1.2 \times 10^{-2}$ &$7.4 \times 10^{-4}$   &$1.3 \times 10^{-2}$ &$1.5 \times 10^{-3}$
\\
\hline
\end{tabular}
}
\end{table*}
%%%%%%%%%%%%%%%%%%%%%%%%%%%%%%%%%%%%%%%%%%%%

\subsection{Fitting using the {\tt BW22} model}
\label{sec:3.2}
The {\tt model 1} fitting procedure described in section~\ref{sec:3.1} is phenomenological in nature, meaning that generic functional forms are fitted, including free parameters that do not necessarily have a direct correspondence to the physical properties of the source. The application of {\tt model 2} was an improvement because the {\tt comptt} component includes a physical temperature. However, even in the case of {\tt model 2}, there is no specification of an overarching physical context. We therefore decided that it would also be interesting to analyze the data using the model of {\tt BW22}, which is designed to simulate the spectral formation process in X-ray pulsars over a wide range of accretion rates. This makes it especially useful for analyzing the outburst spectra for RX J0440.9+4431 in two different luminosity states. The {\tt BW22} model provides a complete description of the spectral formation processes occurring in a conical accretion column located over one of the magnetic poles of an accreting neutron star, and it includes a more realistic velocity profile than that assumed by {\tt BW07}. We briefly review the key elements of the {\tt BW22} model below.

The {\tt BW22} model is a generalization of the original {\tt BW07} model in several respects. These includes the adoption of a conical geometry for the column, which is more appropriate that the cylindrical geometry of the {\tt BW07} model. Furthermore, the {\tt BW22} model also incorporates a more realistic velocity profile that merges smoothly with the Newtonian free-fall profile far from the neutron star. The velocity profiles also incorporates a variable impact velocity, in order to treat cases in the which the accretion rate is sub-critical, and hence radiation pressure is insufficient to decelerate the gas to rest at the stellar surface. In these instances, the final merger with the stellar crust is accomplished via Coulomb deceleration rather than deceleration via the radiation pressure gradient. The {\tt BW22} model also includes the possibility of radiation escape through the walls or top of the accretion column, hence facilitating the computation of ``fan'' and ``pencil'' beam components to the escaping radiation field, respectively. In the {\tt BW22} model, the escape of radiation through the top of the column is treated via the adoption of a free-streaming upper boundary condition. When the accretion rate is super-critical, the deceleration of the gas occurs due to passage through a radiative, radiation-dominated shock wave.

In contrast with a gas-mediated shock, in which the velocity profile exhibits a discontinuity, in a radiation-dominated shock, there is no such discontinuity, and instead the flow velocity decreases smoothly through the sonic point, and the thickness of the shock in this case is about one mean free path for electron scattering \citep{BlandfordAndPayne1981}. When the radiation-dominated shock is ``adiabatic,'' meaning that the total energy flux is conserved, the jump conditions have the standard Rankine-Hugoniot form for a gas with adiabatic index $\gamma = 4/3$, and the velocity decreases by a factor of 7 if the shock is strong \citep{BlandfordAndPayne1981}. However, in an X-ray pulsar accretion column, the shock is radiative in nature, meaning that the total energy flux is not conserved. This is the case treated by {\tt BW07} and {\tt BW22}. In this situation, the kinetic energy is radiated away via photons escaping through the column walls, and the flow essentially comes to rest at the stellar surface \citep{1981Wang,1976Basko,Becker1998}. Hence when radiation escape is taken into account, the compression ratio can be much larger than the factor of 7 obtained in a strong, adiabatic radiation-dominated shock. We note that the smooth, extended region of deceleration in the {\tt BW22} model is qualitatively quite similar to that obtained by \citet{ZhangEtal2022} in their two-dimensional simulations.

Assuming that the source is emitting isotropically at a distance $D = 2.4\,$kpc, we find that the observed X-ray fluxes in the high (ObsID P0514361040) and low (ObsID P0514361005) states correspond to luminosities $L_{\rm X} = 3.44 \times 10^{37}\,{\rm erg\,s}^{-1}$ and $L_{\rm X} = 9.23 \times 10^{36}\,{\rm erg\,s}^{-1}$, respectively. If all of the gravitational potential energy is released before the matter merges with the neutron star crust, and the matter starts out cold and with zero kinetic energy at infinity, then it follows that we can use energy conservation to relate the luminosity, $L_{\rm X}$, to the accretion rate, $\dot M$, via
\begin{equation}
L_{\rm X} = \frac{G M_* \dot M}{R_*} \ ,
\label{eq:lum}
\end{equation}
where $G$ denotes the gravitational constant. Adopting canonical values for the neutron star mass $M_* = 1.4\,M_\odot$ and radius $R_* = 10\,$km, we obtain for the accretion rates in the high (ObsID P0514361040) and low (ObsID P0514361005) states $\dot M = 1.90 \times 10^{17}\,{\rm g\,s}^{-1}$ and $\dot M = 5.00 \times 10^{16}\,{\rm g\,s}^{-1}$, respectively.

In Fig.~\ref{fig:BWspec}, we present the qualitative best fit results for the theoretical phase-averaged spectra obtained in the high and low states using the {\tt BW22} model, compared with the Insight-HXMT data. Individual curves are plotted representing the components of the observed spectrum due to Comptonized bremsstrahlung, cyclotron, and blackbody seed photons, escaping through either the walls or the top of the accretion column. Escape of photons through the top of the column is possible despite the large inflowing mass flux, because the electron scattering cross section for photons propagating parallel to the direction of the magnetic field is about 5 orders of magnitude smaller than the Thomson value (see Tab.~\ref{tab:modpar2}). The values for the associated input theoretical parameters are listed in Tab.~\ref{tab:modpar}. We obtain the same value for the surface magnetic field strength in both states, $B_* = 1.07 \times 10^{13}\,$G, as required since this parameter should remain consistent during the outburst. This is consistent with the estimation by \cite{2024Li} of the magnetic field of RX J0440.9+4431. They considered that the break frequency equals the Keplerian frequency/dynamo frequency at the inner radius of the accretion disk. This value for $B_*$ is used when computing the contribution to the spectrum due to Comptonized cyclotron seed photons.

The cyclotron and bremsstrahlung seed photons are generated throughout the accretion column, but they tend to be concentrated near the base of the column because the gas density is maximized there. The blackbody seed photons are emitted from the ``hot spot'' at the top of the thermal mound, which is located near the base of the accretion column. The spectra plotted in Fig.~\ref{fig:BWspec} indicate that in both the high and low states, the observed phase-averaged spectrum is dominated by Comptonized bremsstrahlung emission, escaping through the walls of the accretion column. The other spectral components make a negligible contribution to the observed spectrum, except the Comptonized cyclotron emission escaping through the column top, which slightly boosts the amplitude of the spectrum around \textasciitilde 100\,keV when the source is in the low state.
We note that the effect of interstellar absorption was not included in the analysis as this is not expected to have a strong effect in the Insight-HXMT energy range.

Here, we are focusing on the simplest case, of the emission from a single magnetic pole, in order to determine how close we can get to the phase-averaged X-ray spectrum observed from RX~J0440.9+4431 during the low and high states of the 2022-2023 giant outburst using the BW22 model. The consideration of emission from only one pole seems to be a reasonable first approach, since it is not clear a priori whether two poles actually contribute to the observed spectrum. Moreover, in order to consider the possible emission from two poles, one would need to employ a complete model for the geometry of the entire star, its rotation, and the alignment of the rotation axis with the line of sight from Earth, as well as specification of the angles between the rotation axis and the magnetic poles, which could may be independent angles if the magnetic field is not exactly dipolar. This general problem is beyond the scope of the present paper, but we plan to consider it in future work.

The {\tt BW22} model is a physics-based analysis tool, and therefore the parameter values obtained using that model can be employed to study the variation of the accretion column structure during the high and low states of the outburst. The values obtained for the theoretical free parameters in the high and low states are reported in Tab.~\ref{tab:modpar}. We note that the electron temperature, $T_e$, has distinctly different values in the high and low states, with $T_e = 1.19 \times 10^8\,$K in the high state and $T_e = 1.40 \times 10^8\,$K in the low state. The lower electron temperature in the high
state results from an increase in the efficiency of bremsstrahlung cooling for larger values of the accretion rate. The parameter $\psi$, which determines the balance between thermal and bulk Comptonization \citep{BeckerWolff2007}, is also significantly different in the low and high states. In the low state, we find that $\psi = 1.50$, and in the high state we obtain $\psi = 1.20$. The lower value of $\psi$ in the high state corresponds to an increase in the level of thermal Comptonization, which tends to increase the number of photons in the energy range \textasciitilde10--20\,keV, corresponding to \textasciitilde $2\,k T_e$. This effect, combined with the lowering of the electron temperature in the high state, may explain the observed flattening of the spectrum as the luminosity approaches the critical luminosity. The spectral flattening may also be connected with changes in the volume and location of the emission region, as discussed in Sec.~\ref{sec:5}.

In addition to the temperature variation noted above, we also find that the radius of the conical accretion column at the stellar surface, $r_0$, also varies significantly over the course of the outburst, with $r_0 = 22.7\,$m in the low state and $r_0 = 43.4\,$m in the high state. These values for the column radius at the stellar surface are comparable to the values obtained by \citet{BeckerWolff2007} and \citet{WolffEtal2016} in their models for Her X-1, which has a similar accretion rate to the source considered here \citep[see also][]{HardingEtal1984}. We interpret the difference between the high-state and low-state column radius as reflecting corresponding variations in the Alfv\'en radius, $R_{\rm A}$, which is the radius in the accretion disk where the accreting matter is redirected by the pulsar's magnetic field \citep{LambEtal1973} into the stream that ultimately connects with the magnetic pole at the stellar surface. The Alfv\'en radius in the disk is computed by balancing the ram pressure of the accreting matter with the magnetic pressure in the dipole field. Within this radius, the fully-ionized gas becomes entrained by the magnetic field, and it is constrained to follow the dipole field lines down to the stellar surface \citep{HardingEtal1984}.

The evolution of the Alfv\'en radius during the outburst depends on the details of the complex interaction between the neutron star's magnetosphere and the accretion disk, which are beyond the scope of the model considered here. However, we note that in general, we expect that in the high state, the greater accretion rate increases the ram pressure, which forces the accreting gas deeper into the dipole field before the field strength becomes large enough to balance the ram pressure of the gas. Hence this effect tends to decrease $R_{\rm A}$ and increase $r_0$ going from the high to the low state, in agreement with the behaviour observed here \citep{LambEtal1973}.

The theoretical spectra plotted in Fig.~\ref{fig:BWspec} also include the effect of cyclotron absorption, with the shape of the CRSF simulated via multiplication by a Gaussian feature centered at the cyclotron energy, $\epsilon_c$, and given by \citep[e.g.,][]{HeindlChakra1999,OrlandiniEtal1998,SoongEtal1990}
\begin{equation}
A_c(\epsilon) \equiv 1 - \frac{d_c}{\sigma_c
\sqrt{2 \pi}}
\, e^{-(\epsilon-\epsilon_c)^2
/(2\sigma^2_c)}
\ ,
\label{eq:Gauss}
\end{equation}
where $\sigma_c$ and $d_c$ represent the width and strength parameters, respectively, and $\epsilon_c$ denotes the cyclotron energy at the imprinting radius, $r_c$, which is the radius where the effect of the CRSF absorption is concentrated. The value of $\epsilon_c$ is related to the field strength, $B_c$, at the imprinting radius $r_c$ via
\begin{equation}
\epsilon_c = 11.57 \, \left(\frac{B_c}{10^{12}\,{\rm G}}\right) \ {\rm keV} \ ,
\label{eq:epsB}
\end{equation}
and the surface magnetic field strength, $B_*$, is related to the CRSF imprinting radius field strength, $B_c$, via the dipole expression
\begin{equation}
    B_c = B_* \left(\frac{r_c}{R_*}\right)^{-3} \ .
\end{equation}

In addition to the input theoretical parameters listed in Tab.~\ref{tab:modpar}, we have also computed several diagnostic output quantities using various theoretical relations from {\tt BW22}, with values reported in Tab.~\ref{tab:modpar2}. These quantities include the scattering cross sections for photons propagating parallel or perpendicular to the magnetic field, denoted by $\sigma_{||}$ and $\sigma_\perp$, respectively. The values obtained for the cyclotron absorption parameters are reported in Tab.~\ref{tab:modpar3}. In the low state, we obtain $B_c = 4.33 \times 10^{12}\,$G, which implies an imprinting radius $r_c = 1.35\,R_*$. Conversely, the value $B_c = 5.19 \times 10^{12}\,$G obtained in the high state implies an imprinting radius $r_c = 1.27\,R_*$. These results indicate that the CRSF is imprinted at an altitude \textasciitilde$0.3\,R_*$ above the stellar surface, which is consistent with the findings of \citet{FerrignoEtal2011} in their study of 4U~0115+63, a source possessing a similar accretion rate to that of RX~J0440.9+4431 during the 2022-2023 giant outburst. Furthermore, we note that the vertical extent of the emission region we find in RX~J0440.9+4431 is comparable to that obtained by \citet{ZhangEtal2022} and \cite{ShengEtal2023} in their two-dimensional simulations of the dynamics of sources with similar accretion rates.

Comparing the high-state and low-state results for cyclotron imprinting radius, $r_c$, indicates that the CRSF is formed at a lower altitude above the stellar surface in the high state as compared to the low state. This is consistent with the analysis presented by \cite{2012Becker}, who found that for sources with luminosity $L_{\rm X} \lesssim L_{\rm crit}$, the emission height tends to decrease with increasing luminosity, where $L_{\rm crit}$ is the critical luminosity for radiation pressure to decelerate the flow.
The width parameter, $\sigma_c$, represents a combination of the intrinsic thermal and natural line broadening, plus the effect of the field gradient through the column height, which causes the cyclotron energy to vary over a large range, changing by a factor of \textasciitilde 30 between the stellar surface and the top of the column, located at radius $r_{\rm top} = 3.1\,R_*$.

It is interesting to examine the results obtained for the various scattering cross sections in the low and high states. In keeping with the expectation of Thomson scattering in the direction perpendicular to the magnetic field in luminous sources, we set $\sigma_\perp/\sigma_{_{\rm T}} = 1$ in both the low and high states \citep{BeckerWolff2007,1981Wang}. On the other hand, for photons propagating parallel to the magnetic field, we find that the scattering cross section is $\sigma_{||}/\sigma_{_{\rm T}} = 1.23 \times 10^{-5}$ in the high state, and $\sigma_{||}/\sigma_{_{\rm T}} = 1.89 \times 10^{-5}$ in the low state. These values are quite similar, and the minor difference between them can be attributed to the different values for the CRSF energy, $\epsilon_c$, obtained in the two states, with $\epsilon_c = 50.1\,$keV in the low state, and $\epsilon_c = 60.0\,$keV in the high state. The lower value of $\epsilon_c$ in the low state implies that the continuum energy region is closer to the the cyclotron resonance, which tends to amplify the scattering cross section \citep{1981Wang}.

It is also interesting to examine the variation of the dynamical structure of the accretion column during the outburst. In both the high and low states, we find that the dynamical parameter $k_\infty = 1$, indicating that the velocity profile of the accreting gas merges smoothly with Newtonian free-fall far from the neutron star, as expected. However, the dynamical parameter $k_0$, which determines the impact velocity of the gas at the stellar surface, is different in the two states. In the high state, we find that $k_0 = 0$, indicating that the gas settles onto the neutron star surface with zero velocity, which is the expected behaviour in high-luminosity accretion-powered X-ray pulsars due to the effect of the strong radiation pressure gradient force in these sources \citep{Davidson1973,1976Basko}.

In the low state, we find that $k_0 = 0.1$, which indicates that the gas collides with the stellar surface with impact velocity $\upsilon_* = 0.064\,$c. This occurs because the radiation pressure gradient force is too small to entirely decelerate the accreting gas to rest at the stellar surface. In this case, the final deceleration and merger of the accreting matter with the neutron star crust presumably occurs via passage through a thin discontinuous shock, and the residual kinetic energy in the accretion flow powers the emission of the blackbody seed photons from the hot spot at the stellar surface \citep{LangerAndRappaport1982,BeckerAndWolff2005a,BeckerAndWolff2005b}.

 In the {\tt BW22} model, the velocity profile defined by the constants $k_0$ and $k_\infty$ is considered to be an input to the model, and the corresponding emergent spectrum is computed based on this input velocity profile, combined with a realistic radiation transport equation and a set of appropriate physical boundary conditions. Although the velocity profile is considered to be an input to the model, its validity can be evaluated after the fact by using the resulting radiation spectrum and gas temperature and density variations to compute the pressure gradients, which can be combined with the gravitational force to re-compute the velocity profile by solving numerically the hydrodynamical conservation equations for the gas. This procedure was carried out by {\tt BW22} for their model of Her X-1 and the velocity profile was found to be essentially physically self-consistent, which is also the case here.

\section{Discussion}
\label{sec:4}

The Type II giant outburst of RX~J0440.9+4431 in 2022--2023 was intensively observed using Insight-HXMT. By analyzing the the wide-band Insight-HXMT spectra during the outburst, we find that as the luminosity increases, the spectrum of this source gradually changes. In particular, when the luminosity exceeds 3$\times10^{37}\ {\rm erg\ \rm s^{-1}}$, the emission increases significantly in the energy ranges \textasciitilde 2\,keV and \textasciitilde 20--40\,keV. This spectral evolution causes the SC to decrease as the count rate increases when the source is bright. Conversely, the HC increases continually as the count rate increases (see Fig.~\ref{fig:4}). This behavior is consistent with the work of \cite{2013Reig}, who conducted a systematic study on HID/SC of Be X-ray binaries and identified two spectral branches: the horizontal branch corresponds to a low-intensity state of the source, and the diagonal branch represents a high-intensity state occurring when the X-ray luminosity exceeds a critical limit. Therefore, they concluded that the spectrum softens in the high state for many sources, in agreement with our findings during the outburst of RX~J0440.9+4431 studied here.

However, if we focus on higher energies, we note that the HC of RX~J0440.9+4431 increases as the count rate rises, indicating that the high-energy spectrum becomes harder. We note that the HC of KS 1947+300, 4U 0115+63, and EXO 2030+375 in \cite{2008Reig} also exhibits a similar trend. Therefore, we cannot solely rely on the changes in SC to claim that the spectrum softens, and these variations in SC and HC are likely common characteristics of the majority of Be X-ray binaries during the outburst. Fig.~\ref{fig:1} aids in better understanding these two features and also suggests that Be X-ray binaries such as KS 1947+300, 4U 0115+63, and EXO 2030+375 may undergo similar changes to RX~J0440.9+4431 in their high states, especially with increases in the intensity for photon energies \textasciitilde 2\,keV and \textasciitilde 20--40\,keV. Hence it may be more accurate to describe the spectral evolution as a flattening of the spectrum in the high state, which is apparent in the spectra plotted in Fig.~\ref{fig:2}. 
Additionally, this behavior, as seen from the right panel of Fig.~\ref{fig:1}, is phase-dependent, implying a direct association with radiation from the accretion column.

We selected two ObsIDs from the Insight-HXMT data to represent the low and high states of RX J0440.9+4431. Fig.~\ref{fig:2} shows the results fitted with commonly used phenomenological models. As mentioned earlier, there is a significant increase in emission around \textasciitilde 2\,keV and \textasciitilde 20--40\,keV in the high state. This naturally implies the need for a low-temperature blackbody at high states, while another blackbody, to account for the emission at high energies, becomes very hot. Additionally, the {\tt comptt} model requires a larger optical depth to achieve a flatter spectrum. While these models can fit the spectra well, gaining detailed insight into the fundamental accretion physics remains challenging.

To further study the radiation mechanism of RX J0440.9+4431, we also used the {\tt BW22} model to qualitatively fit the spectra, since this model provides a complete physical description of the spectral formation processes occurring in an X-ray pulsar accretion column. The relative radiated power distributions over the emission components obtained using the {\tt BW22} model in the high and low states are reported in Tab.~\ref{tab:lumdist}. We note that in both the high and low states, the radiated power is dominated by Comptonized bremsstrahlung emission, with \textasciitilde 90\% of this emission escaping through the column walls. However, the proportion of the emission escaping through the column top decreases in the low state. This can be understood as follows. In the low state, due to the dynamical parameter value $k_0 = 0.1$, the average speed of the accretion flow is higher than that it is in the high state, with $k_0 = 0$. The increased flow velocity in the low state enhances the level of photon ``trapping,'' so that the radiation tends to be swept downstream, resulting in a higher proportion of the emission escaping through the column walls, rather than the top, when the source is in the low state \citep{BeckerWolff2007}.

The plot of the pulse profile on the right-hand side of Fig.~\ref{fig:1} indicates that when the source is in the high state, the spectrum becomes significantly harder near the peak of the pulse profile. It is interesting to correlate this with the plot of the phase-averaged spectrum computed using the {\tt BW22} model, depicted in the right-hand panel of Fig.~\ref{fig:BWspec}. The theoretical results for the spectrum show that in the high state, the bremsstrahlung component emitted from the column top is significantly harder than that emitted through the column walls. However, the column top component makes a weaker contribution to the phase-averaged spectrum than does the wall component. Hence the question of whether the hardening of the spectrum at the peak of the pulse can be explained by the {\tt BW22} model depends on the details of the beaming pattern for the radiation escaping through the column top, which we briefly discuss below.

The emission escaping through the column top is expected to be strongly beamed in the direction of the magnetic field, which is essentially the radial direction. This focusing effect will tend to boost the contribution of the column top emission to the pulse profile, if the column top is aligned with the distant observer at the peak of the pulse. Hence in this scenario, the focused emission from the column top could produce a significant hardening of the spectrum at the pulse peak, as observed. While this is an interesting possibility, ultimately, this question can't be answered in a quantitative way without a phase-resolved calculation of the theoretical spectral components escaping through the walls and top of the accretion column, which is beyond the scope of this paper. We plan to compute phase-resolved theoretical spectra in future work.

It is instructive to compare the behavior of RX J0440.9+4431 with that of Her X-1, which was treated by \cite{BeckerWolff2022} using the same model employed here. We note that the accretion rate for Her X-1 is similar to that in the high state of RX~J0440.9+4431. Based on its magnetic field strength, derived from CRSF observations, Her X-1 is estimated to be super-critical, with $L_{\rm X} \gtrsim L_{\rm crit}$, when at maximum luminosity \citep{2012Becker}. On the other hand, because of its stronger estimated magnetic field strength, RX~J0440.9+4431 may be quasi-critical when in the highest state considered here, with $L_{\rm X} \sim L_{\rm crit}$. The behaviors of the two sourecs are similar, although there are some differences. For example, the spectrum of Her X-1 tends to get harder with increasing luminosity \citep{StaubertEtal2020,StaubertEtal2019}, and RX~J0440.9+4431 likewise exhibits an increase in high-energy radiation with increasing luminosity, although the spectrum also becomes flatter in the energy range around \textasciitilde 2\,keV and 20--40\,keV.
Furthermore, in both Her X-1 and RX~J0440.9+4431, we find that the radiation emerges primarily from the side of the accretion column rather than from the top, which is consistent with expectations for sources in this luminosity range \citep{BeckerWolff2007}. The slight differences in behavior may stem from the stronger magnetic field obtained in the case of RX~J0440.9+4431, which leads to electron scattering cross sections that are a factor of \textasciitilde 3 smaller than those obtained in the case of Her X-1 \citep{BeckerWolff2022}.

\section{Conclusion}
\label{sec:5}

We have analyzed the evolution of the X-ray spectrum from the BeXRB source RX~J0440.9+4431 during the Type~II giant outburst that occurred in 2022--2023. We have interpreted the data via phenomenological fitting, and also via comparisons with simulations carried out using the {\tt BW22} physics-based model. One important result of the phenomenological fitting is that as the luminosity approached the critical luminosity during the outburst, the spectrum became flatter, with the increase in the spectrum primarily concentrated around \textasciitilde 2\,keV, with an accompanying hardening in the energy range 20--40\,keV. This behavior is consistent with the results of \cite{2011Reig}, who conducted a study of several variable BeXRB systems, and found that as the source reaches its highest luminosity, the SC decreases (along the diagonal branch), but the HC increases, indicating hardening at higher photons energies. We obtain similar results here, as indicated in Fig.~\ref{fig:4}.

In order to obtain additional physical insight, we also performed simulations using the {\tt BW22} theoretical model. The results indicate that the electron temperature in the accretion column was $T_e = 1.19 \times 10^8\,$K in the high state and $T_e = 1.40 \times 10^8\,$K in the low state. The lower electron temperature in the high state is driven by an increase in the efficiency of bremsstrahlung cooling at higher accretion rates. Furthermore, we find that the Comptonization parameter $\psi$ decreases in the high state, indicating a decrease in bulk Comptonization relative to thermal Comptonization. These factors combine to explain the flattening of the spectrum in the high state. We also find that the CRSF formation radius decreases in the high state. Taken together, these facts provide the first quantitative explanation for the suggestion made by \cite{2013Reig} that the spectral softening in the high state is a consequence of a decrease in the altitude above the stellar surface where most of the emission is produced, leading to a compactification of the emission region, which allows the photons to diffuse more effectively out of the region of maximum bulk Comptonization.

%__________________________________________________________________
\begin{acknowledgements}
We thank the anonymous referee for valuable comments and suggestions.

We also thank the Insight-HXMT science team for the valuable suggestions, which have improved the paper. This research utilized data and software from the Insight-HXMT mission, supported by the China National Space Administration (CNSA) and the Chinese Academy of Sciences (CAS).
Financial support for this work is provided by the National Key R\&D Program of China (2021YFA0718500). We also acknowledge funding from the National Natural Science Foundation of China (NSFC) under grant numbers 12122306, 12333007, and U2038102, as well as support from the CAS Pioneer Hundred Talent Program (Y8291130K2), the Strategic Priority Research Program of the Chinese Academy of Sciences (No. XDB0550300), and the Scientific and Technological Innovation Project of IHEP (Y7515570U1).

\end{acknowledgements}

%-------------------------------------------------------------------

\bibliographystyle{aa}
\bibliography{ref}
\end{document}